\documentstyle[11pt,epsfig]{article}
\newlength{\bredde}
\def\slash#1{\settowidth{\bredde}{$#1$}\ifmmode\,\raisebox{.15ex}{/}
\hspace*{-\bredde} #1\else$\,\raisebox{.15ex}{/}\hspace*{-\bredde} #1$\fi}
\newcommand{\beq}{\begin{equation}}
\newcommand{\eeq}{\end{equation}}
\newcommand{\ba}{\begin{array}{ccc}}
\newcommand{\ea}{\end{array}}
\newcommand{\nn}{\nonumber}
\newcommand{\noi}{\vspace{12pt}\noindent}
\renewcommand{\d}{\partial}
\def\beqn{\begin{eqnarray}}
\def\eeqn{\end{eqnarray}}

\def\al{\alpha}

\def\psit{\tilde{\psi}}
\def\Tr{ {\rm Tr} }
\def\<{\langle}
\def\>{\rangle}
\begin{document}
\topmargin -1.4cm
\oddsidemargin -0.8cm
\evensidemargin -0.8cm

\title{\Large{{\bf QCD-like Theories \\ at Finite Baryon and Isospin Density}}}
\author{~{\sc  K. Splittorff$^1$}, {\sc D.T.Son$^{2,4}$},  and  {\sc
    M.A. Stephanov$^{3,4}$} 
\\ ~ \\
\footnotesize{$^1$ The Niels Bohr Institute, Blegdamsvej 17, DK-2100 Copenhagen
  {\O}, Denmark} \\
\footnotesize{$^2$ Physics Department, Columbia University, New York, NY 10027, USA} \\
\footnotesize{$^3$ Department of Physics, University of Illinois, Chicago, IL 
60607-7059, USA} \\
\footnotesize{$^4$ RIKEN-BNL Research Center, Brookhaven National Laboratory,
Upton, NY 11973, USA} }
\date{\today }
\maketitle

\vfill
\begin{abstract} 
We use 2-color QCD as a model to study the effects of simultaneous
presence of chemical potentials for isospin charge, $\mu_I$, 
and for baryon number, $\mu_B$.
We determine the phase diagrams for 2 and 4 flavor theories
using the method of effective chiral Lagrangians at low densities
and weak coupling perturbation theory at high densities. We determine
the values of various condensates and densities as well as the spectrum of
excitations as functions of $\mu_I$ and $\mu_B$. A similar analysis of QCD
with quarks in the adjoint representation is also presented. Our results
can be of relevance for lattice simulations of these theories.
We predict a phase of inhomogeneous condensation
(Fulde-Ferrel-Larkin-Ovchinnikov phase) in the 2 colour 2 flavor
theory, while we do not expect it the 4 flavor case or in other realizations of 
QCD with a positive measure.  
\end{abstract}

\vfill

%\begin{flushleft}
%NBI-NT-??-?? \\
%hep-ph/0012274
%\end{flushleft}
\thispagestyle{empty}
\newpage

\section{Introduction}

 QCD at non-zero chemical potential has been a subject of numerous
studies recently. The phenomenon of quark-quark pairing and
color-superconductivity \cite{BL} has received renewed attention in
view of the recent observations \cite{ARF} that the superconducting gaps
can be large and of possible relevance in astrophysics and
heavy ion collisions (see \cite{RW} for a review). The most interesting finite density
phenomena (phase transitions, in particular) occur in the
non-perturbative regime, thus inviting lattice calculation
methods. However, the lack of positivity of the Euclidean
path integral prevents straightforward application of lattice
Monte Carlo techniques. QCD at finite density thus remains
a theoretical challenge. \\
This is one of the reasons that certain QCD-like theories with
positive Euclidean path integral measure have also attracted
attention.  Examples of such theories are 2-color QCD, QCD
with adjoint quarks, 
or 2-flavor QCD with isospin chemical
potential \cite{kogut1,kogut2,Hands,AKW,SS}. 
In these theories, for each quark with a chemical
potential $\mu$ there is another (conjugate) quark with exactly the
same properties, except for the opposite sign of $\mu$.  The quenched
approximation, common in lattice studies, is an approximation to such
kind of theories, rather than to QCD at finite baryon density with all
quarks having equal chemical potentials \cite{St96}.

In reality, dense baryon or quark matter, such as that arising in the
interior of neutron stars or heavy ion collisions, is characterized by
different, not equal and not opposite, chemical potentials for
different flavors of quarks. One can describe such situations by the
simultaneous presence of chemical potentials for baryon charge,
$\mu_B$, and isospin $\mu_I$. The physics of this has been
discussed \cite{ABR} in the context of neutron stars ($\mu_B\gg\mu_I$)
and QCD at large density of isospin and small baryon density 
($\mu_I\gg\mu_B$) \cite{SS}. The most interesting result is the appearance
of a phase where ground state breaks translational and rotational
symmetry. This phase is similar to the so-called FFLO
(Fulde-Ferrel-Larkin-Ovchinnikov) phase of a BCS superconductor
\cite{FFLO}.

In this paper we investigate the behavior of  QCD-like theories, 
in particular 2-color QCD under the influence of both $\mu_I$ and $\mu_B$.
We shall use controllable analytical methods in two regimes
of these theories: a chiral effective Lagrangian, based on the
symmetries, in the low density regime, and weak coupling perturbation theory in 
the high density regime. We shall determine the phase diagram in the
$\mu_B,\mu_I$ plane, various condensates, and lowest lying excitations.
 
Our goal is to gain an understanding of the physics of the interplay
between $\mu_B$ and $\mu_I$. The theories we study (except
$N_c=N_f=2$) have the advantage of having positive Euclidean path integral 
measure, and can, therefore, be studied on the lattice. Our results
can be used as benchmarks for numerical lattice studies.

\noi
The presentation of our results is organized as follows. We start 
with  two-colour QCD. In section
\ref{SectL} we give the low energy effective Lagrangian upon which the 
sequel is based.  A qualitative argument for the structure of the phase
diagram is then given in \ref{sec:qualitative}.
In section \ref{SectClassical} we analyse the properties of
the vacuum. As a result we find the classical values for the condensates and 
densities. Then in  section \ref{SectQuantum} we study the expansion about
the minimum in order to determine the masses of the low energy
excitations. The extension of this theory to $N_f=4$ with two up and two down
flavours is discussed in section \ref{sec:nf=4}. The discussion of 
asymptotically high chemical potentials and the FFLO-phase is in section 
\ref{sec:FFLO}.
Finally, in the last section we summarize and comment on the relevance
of our results for real QCD. The presentation of QCD with quarks in the adjoint
representation is placed in the appendix.

\section{The Effective Theory at Finite $\mu_B$ and $\mu_I$}
\label{SectL}

Two colour QCD at zero chemical potential is invariant under SU(2$N_f$)
rotations in the chiral limit, see e.g \cite{SV}. This enhanced symmetry
(as compared to the
SU($N_f)\times$SU($N_f)\times$U(1) of three colour QCD)  is manifest 
in the Lagrangian if we choose to represent this in a basis of quarks $\psi$
and conjugate quarks $\psit$ \cite{kogut1}. For $N_f=2$ we use
\beq
\Psi\equiv
\left(\begin{array}{c} u \\ d \\ \tilde u \\ \tilde d
  \end{array}\right)\equiv\left(\begin{array}{c} u_L \\ d_L \\
    \sigma_2\tau_2(u_R)^* \\ \sigma_2\tau_2(d_R)^*
  \end{array}\right)
\label{basis}
\eeq 
% \beq
% \left(\begin{array}{c} \psi_L^u \\ \psi_L^d \\ \psit_R^u \\ \psit_R^d
%   \end{array}\right)\equiv\left(\begin{array}{c} \psi_L^u \\ \psi_L^d \\
%     \sigma_2\tau_2(\psi_R^u)^* \\ \sigma_2\tau_2(\psi_R^d)^*
%   \end{array}\right)
% \label{basis}
% \eeq 
where the Pauli matrices $\tau_2$ and $\sigma_2$ act in colour and spin space respectively.

The enhanced symmetry manifests itself in the low energy effective theory
through the manifold of the Goldstone modes associated with the spontaneous
breaking of chiral symmetry. In our case $N_f = 2$ and the Goldstone
manifold is SU(4)/Sp(4), corresponding to the condensation
of $\Psi\Psi$ --- SU(4) flavor sextet. The fields on this manifold can be
represented by a  4$\times4$ antisymmetric unitary matrix $\Sigma$
with $\det\Sigma=1$.

The effective Lagrangian for the field $\Sigma$ of Goldstone modes is
determined by the symmetries inherited from the microscopic two-colour QCD
Lagrangian. 
%The restriction of the effective Lagrangian for different 
%chemical potentials is completely analogous to that for baryon chemical
%potential studied in \cite{kogut1,kogut2}. 
In the basis of SU(4) spinors (\ref{basis}) the mass
matrix, baryon charge matrix, 
and the isospin (third component) charge matrix are%
\footnote{Note on the notations: 
We use the normalization of the baryon charge different from
\cite{kogut2}. The baryon charge of the quark is not 1, as in
\cite{kogut2}, but 1/2, which comes from $1/N_c$, so that
the baryon (diquark in $N_c=2$) has baryon charge 1. For simplicity,
we omit subscript $3$ from $I_3$.
}
\beqn
\label{MBandI}
{\cal M}\equiv\left(\begin{array}{cccc} 0 & 0   & 1 & 0 \\  0 & 0  & 0 & 1\\
    -1 & 0  & 0  & 0\\ 0 & -1  & 0  & 0\\\end{array}\right) ,\ 
B\equiv \frac12
\left(\begin{array}{cccc} 1 & 0  & 0  & 0\\ 0 & 1  & 0  & 0\\ 0 & 0  &
    -1  & 0\\ 0 & 0  & 0  & -1\\ \end{array}\right) ,\ 
I\equiv \frac12
\left(\begin{array}{cccc} 1 &  0 &  0 & 0 \\  0 & -1 & 0  & 0\\ 0 & 0  & -1  & 0\\ 0 & 0 & 0 & 1\\   \end{array}\right)
\eeqn
and the resulting effective Lagrangian is 
\beqn
\label{L}
{\cal L}_{\rm eff}(\Sigma) &=&  {F^2\over2} \Tr \d_\nu \Sigma \d_\nu \Sigma^\dagger 
+ 2 F^2\Tr ((\mu_BB+\mu_II) \Sigma^\dagger\d_0 \Sigma)
\nonumber\\&&
%\hskip 1em
 -F^2\Tr\left(\Sigma(\mu_BB+\mu_II) \Sigma^\dagger (\mu_BB+\mu_II)+
   (\mu_BB+\mu_II)^2\right)-F^2m_\pi^2{\rm Re}\Tr\left({\cal M}\Sigma\right)
 \ . 
\eeqn
The $\mu$-dependent terms in the effective Lagrangian
appear through the covariant extension of the derivative:
\beqn
 \partial_0 \Sigma & \to & \partial_0 \Sigma -
\left[(\mu_BB+\mu_II)  \Sigma +  \Sigma (\mu_BB+\mu_II)^T\right] \\
 \partial_0 \Sigma^\dagger & \to & \partial_0 \Sigma +
\left[(\mu_BB+\mu_II)  \Sigma +  \Sigma (\mu_BB+\mu_II)^T\right]^\dagger
\eeqn
required by an extended local gauge symmetry \cite{kogut1}.
Therefore, to this order in chiral perturbation theory, 
the Lagrangian at finite $\mu$ does not require
any extra phenomenological parameters beyond the pion decay constant, $F$, 
and the chiral condensate in
the chiral limit, $G$. This fact gives predictive power
to chiral perturbation theory at finite $\mu$.
The chiral condensate can be traded for the vacuum pion mass
using the Gell-Mann-Oaks-Renner relation ($m=m_u=m_d$ is the quark mass): 
\beq
F^2m_\pi^2=mG \ .
\eeq

In using the effective Lagrangian constructed above we must of course assume 
that chiral symmetry for $N_c=N_f=2$ QCD is spontaneously broken.
 Since we have
regarded the hadronic modes as heavy the theory is expected to be valid only
up to the mass of the lightest non-goldstone hadron. However, as we shall show the phase diagram is very rich in the effective range of the theory.

\section{The phase diagram expected from the $\mu_I=0$ spectrum}
\label{sec:qualitative}

One can understand main features of the phase
diagram in the $\mu_B\mu_I$ plane 
by considering the low-energy spectrum of the theory
at finite $\mu_B$ determined in \cite{kogut2}.
In the vacuum, $\mu_B=\mu_I=0$, the spectrum consists
of a degenerate 5-plet: 3 pions $\pi_0$, $\pi^\pm$ 
($B=0$, $I=0,\pm1$), a  baryon and an antibaryon $q$, $q^*$
(diquark and antidiquark with $B=\pm1$ and $I=0$).

On the horizontal axis, $\mu_I=0$, as a function of $\mu_B$,
there is a phase
transition corresponding to the condensation of diquarks.
This happens at $\mu_B$ equal to the diquark mass divided
by its baryon charge, i.e., at $\mu_B=m_\pi$. Similarly,
at $\mu_B=0$ there should be a transition corresponding
to the condensation of pions. This happens when $\mu_I$
is equal to the pion mass divided by its isospin charge, i.e.,
at $\mu_I=m_\pi$. This can also be concluded from the fact that
there is a simple discrete symmetry: $d \leftrightarrow \tilde d$
accompanied by $\mu_B\leftrightarrow\mu_I$. Therefore, the $\mu_B\mu_I$ phase
diagram must be symmetric under reflection 
about $\mu_B=\mu_I$ line. 

As a function of $\mu_B$ the mass of the pion  is constant as long as
$\mu_B<m_\pi$,
which means the transition at $\mu_I=m_\pi$ happens at 
all such $\mu_B$. Related to this horizontal line of phase transitions
by a reflection against the diagonal, there is a vertical line at
$\mu_B=m_\pi$, the line where diquarks condense. After the
two lines meet they must merge into a single $\mu_B=\mu_I$
line. This is because of the reflection symmetry and the fact
that the mass of the pion in the diquark
condensation phase $\mu_B>m_\pi$ is equal to 
$\mu_B$ according to \cite{kogut2}.
Two different condensation modes (pion vs diquark) compete 
at $\mu_B=\mu_I$, thus the transition
along the $\mu_B=\mu_I$ line is of first order.

\section{Minimum of the effective Lagrangian}
\label{SectClassical}

\noi
In order to derive the phase diagram we study the minimum of the classical
theory.  
That is: we seek to minimize the static part of the effective Lagrangian
\beq
\label{Lst}
{\cal L}_{\rm eff}(\overline{\Sigma}) ~ = ~   -F^2
\Tr\left(\overline{\Sigma}(\mu_BB+\mu_II) \overline{\Sigma}^\dagger
  (\mu_BB+\mu_II)+ (\mu_BB+\mu_II)^2\right) 
-F^2m_\pi^2{\rm Re}\Tr\left({\cal M}\overline{\Sigma}\right) 
\eeq
of the Lagrangian (\ref{L}). Our ansatz for the minimum is
\beq
\overline{\Sigma}\equiv
\Sigma_M\cos\al+(\Sigma_B\cos\eta+\Sigma_I\sin\eta)\sin\al \ , 
\label{ansatz}
\eeq
where $\al$ and $\eta$ are variational parameters, $\Sigma_M\equiv-{\cal M}$, while 
$\Sigma_B$ and $\Sigma_I$ are given by 
\beq
\Sigma_B = \left(\begin{array}{cccc}
0 & -i & 0 & 0 \\
i & 0 & 0 & 0 \\
0 & 0 & 0 & -i \\
0  & 0 & i & 0 \\
\end{array}\right) \ \ \ {\rm and} \ \ \ \ 
\Sigma_I = \left(\begin{array}{cccc}
0 & 0 & 0 & i \\
0 & 0 & i & 0 \\
0 & -i & 0 & 0 \\
-i & 0 & 0 & 0 \\
\end{array}\right) \ .
\label{Sigma_d}
\eeq
The ansatz is based on the following 
observations: three terms compete for the alignment of the
condensate: $m{\cal M}$, $\mu_BB$, and $\mu_II$. Each of them is
independently minimized by  $\Sigma_M$, $\Sigma_B$, and
$\Sigma_I$ respectively.

Inserting the ansatz and using the (anti)commutation relations between the
matrices involved we reduce the static Lagrangian to
\beq
{\cal L}_{\rm eff}(\overline{\Sigma})~=~-  F^2 \left(-(\mu_B^2+\mu_I^2)\cos^2\al+(\mu_B^2-\mu_I^2)\cos(2\eta)\sin^2\al+ \mu_B^2+\mu_I^2+4m_\pi^2\cos\al\right) \ . 
\label{LstSigma}
\eeq
Extremizing with respect to $\eta$ we find (for $\alpha\ne0$):
\beqn
\eta = 0 & {\rm if} & \mu_I<\mu_B\nn\\
\eta = \frac\pi2 & {\rm if} & \mu_I>\mu_B  \ .
\eeqn
Note that for $\mu^2_B=\mu^2_I$ there is degeneracy in the $\eta$-direction. 
 This is a manifestation of the SU(2) rotational symmetry (at $m=0$) between
 the flavor which has $\mu=0$ and its own conjugate field (i.e., between $d$ and
$\tilde{d}$ in the first quadrant of the $\mu_I,\mu_B$-plane).   
Extremizing with respect to the $\al$-direction we find:
\beqn
\al = 0 & {\rm if} & y > 1\nn \\
\cos\al = y & {\rm if} & y < 1 
\label{al}
\eeqn
 where
\beqn
y = {m_\pi^2\over\mu_B^2} & {\rm if} & \mu_I<\mu_B\nn\\
y = {m_\pi^2\over\mu_I^2} & {\rm if} & \mu_I>\mu_B \ .
\label{y}
\eeqn

The ansatz (\ref{ansatz}) is indeed a minimum of the classical theory as we
shall prove in the next section. 

\begin{figure}[t]
\epsfig{file=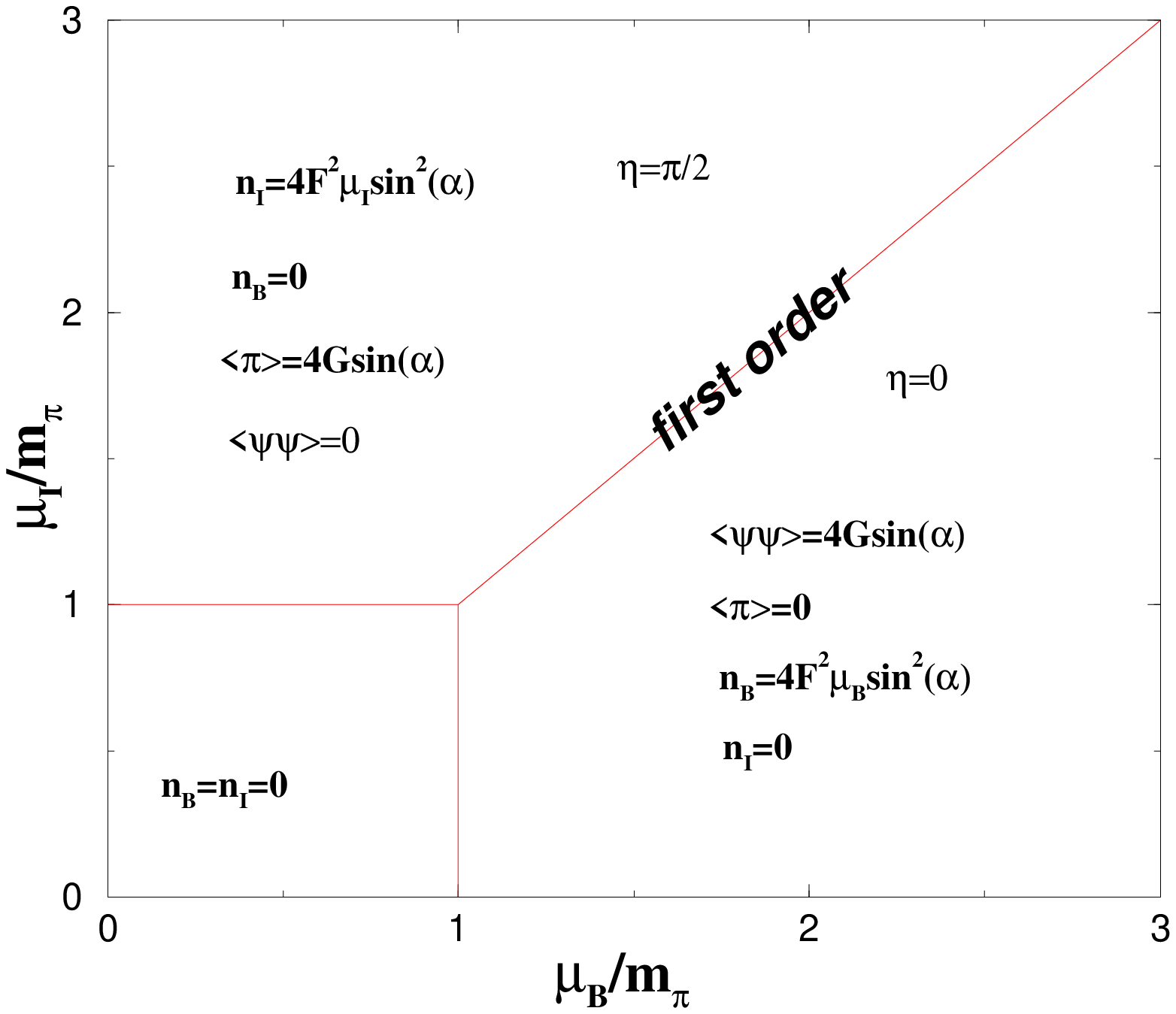,clip=,width=7cm}
\hfill 
  \unitlength1.0cm
  \begin{center}
  \begin{picture}(3.0,2.0)
  \put(2.0,-3.0){
  \epsfysize=8.5cm
  \epsfbox[110 -180 390 100]{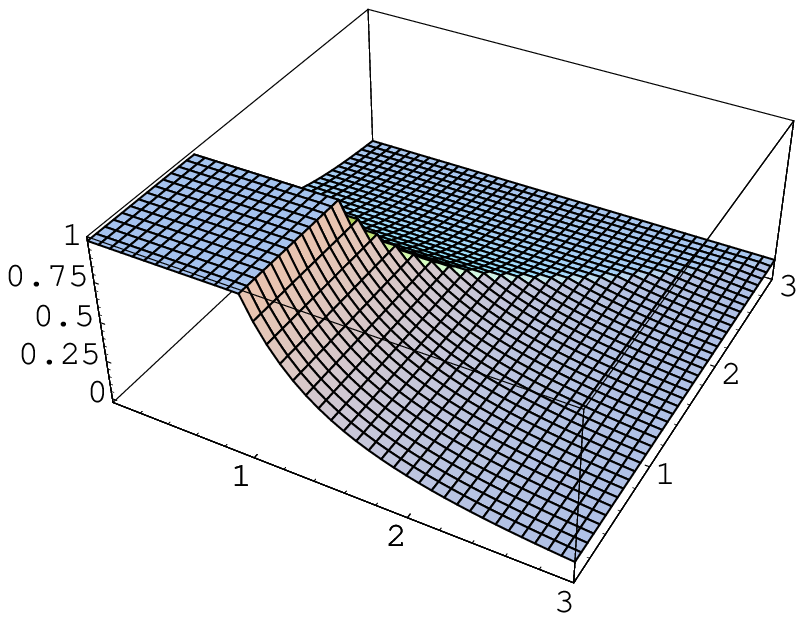}}
  \put(4.0,3.2){\bf\large $\frac{\mu_B}{m_\pi}$}
  \put(8.5,4.2){\bf\large $\frac{\mu_I}{m_\pi}$}
  \put(0.5,5.5){\bf\large $\frac{\langle\bar{\psi}\psi\rangle}{\langle\bar{\psi}\psi\rangle_0}$}
  \end{picture}
  \end{center}
\vspace{-3cm}

\caption{\label{muBmuI-figI} $N_c=2$, $N_f=2$ ($\beta_{\rm D}=1$): 
  The figure on the left shows a schematic version of the
  phase diagram. To the right: The ratio of the chiral condensate to its
  value at $\al=0$ in the first quadrant of the $(\mu_B,\mu_I)$-plane.}
\end{figure}

\begin{figure}[t]
  \unitlength1.0cm
  \begin{center}
  \begin{picture}(3.0,2.0)
  \put(-6.2,-9.5){
  \epsfysize=8.5cm
  \epsfbox[110 -180 390 100]{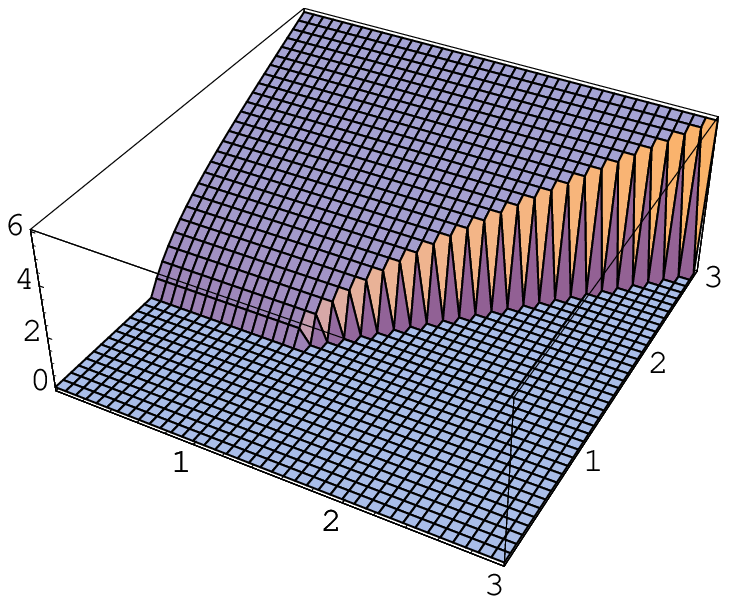}}
  \put(-4.5,-3.2){\bf\large $\frac{\mu_B}{m_\pi}$}
  \put(-1.,-2.2){\bf\large $\frac{\mu_I}{m_\pi}$}
  \put(-7.8,-0.5){\bf\large $\frac{n_I}{2 F^2 m_\pi}$}
  \end{picture}
  \end{center}

\hfill 
  \unitlength1.0cm
  \begin{center}
  \begin{picture}(3.0,2.0)
  \put(2.0,-6.0){
  \epsfysize=8.5cm
  \epsfbox[110 -180 390 100]{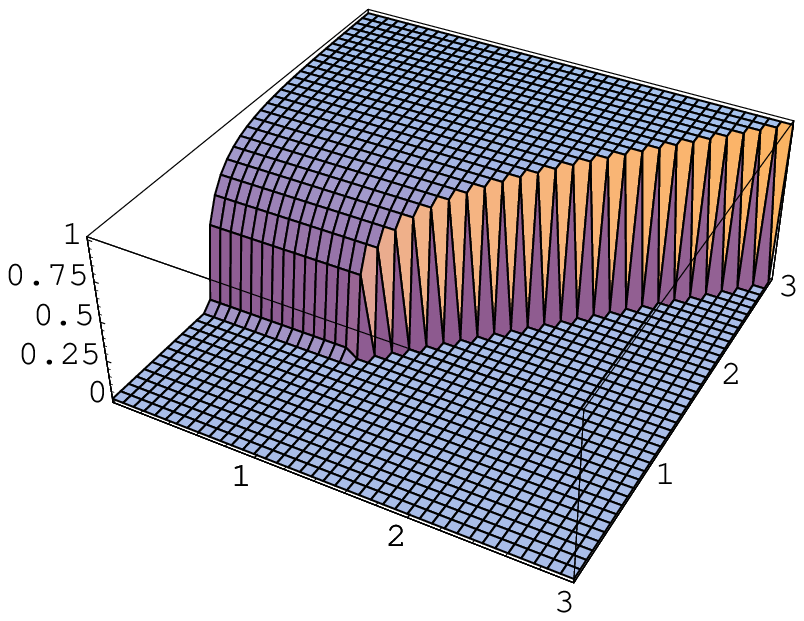}}
  \put(4.0,0.2){\bf\large $\frac{\mu_B}{m_\pi}$}
  \put(8.1,1.2){\bf\large $\frac{\mu_I}{m_\pi}$}
  \put(0.7,2.5){\bf\large $\frac{\langle\pi\rangle}{\langle\bar{\psi}\psi\rangle_0}$}
  \end{picture}
  \end{center}

\caption{\label{muBmuI-figII} $\beta_{\rm D}=1$, $N_f=2$: The leftmost plot displays 
the isospin charge density in the
 $(\mu_B,\mu_I)$-plane in units of $2 F^2 m_\pi$.  On the right hand side is
 shown the ratio of the pion condensate to the chiral condensate at $\al=0$. The first 
order phase transition is apparent.}
\end{figure}

Since the vacuum energy is given by the
value of the Lagrangian at the minimum 
we can draw the following conclusions
\beq
\begin{array}{rll}
 \langle\bar{\psi}\psi\rangle & = ~ -\frac{\d {\cal L}_{\rm eff}(\overline{\Sigma})}{\d
  m}=4G\cos\al  & \nn \\ 
   \nn \\
n_B & =~ -\frac{\d {\cal L}_{\rm eff}(\overline{\Sigma})}{\d \mu_B} =2F^2\mu_B 
(1+\cos(2\eta))\sin^2\al & n_I=-\frac{\d {\cal L}_{\rm eff}(\overline{\Sigma})}{\d \mu_I}
=2F^2\mu_I
(1-\cos(2\eta))\sin^2\al   \\ 
   \nn \\
    \langle\psi\psi\rangle   & = ~ -\frac{\d {\cal L}_{\rm eff}(\overline{\Sigma})}{\d
  j_B}=4G\sin\al\cos\eta & \langle\pi\rangle=-\frac{\d {\cal L}_{\rm eff}(\overline{\Sigma})}{\d
  j_I}=4G\sin\al\sin\eta 
 \ . 
\label{condensates}
\end{array} 
\eeq
To obtain the diquark and pion condensates we have introduced a diquark source, $j_B$,
and a pion source, $j_I$, in (\ref{L}). The condensates are the 
derivatives of ${\cal L}_{\rm eff}(\overline{\Sigma}) $ with respect to the
sources. The introduction of the sources is analogous to  
that in \cite{kogut2} and we refer to that paper for details. The phase diagram is shown in 
the left hand side of fig.\ref{muBmuI-figI}. It is in agreement with our qualitative
argument given in section \ref{sec:qualitative}.
In particular the densities are discontinuous across at $\mu_B=\mu_I>m_\pi$ indicating a first order phase transition. Note, however, that the vacuum energy, Eq. (\ref{LstSigma}), is smooth across the $\mu_B=\mu_I$ line. The values of the condensates are illustrated in figures \ref{muBmuI-figI}
and \ref{muBmuI-figII}.

\section{Expanding About the Minimum}
\label{SectQuantum}

In order to determine the phase diagram we had to find
the minimum of the effective Lagrangian (\ref{L})
with respect to $\Sigma$. Having found this, we now
determine the masses of low-energy excitations
by expanding around the minimum $\overline\Sigma$.

\subsection{Parameterization of the field $\Sigma$}

A convenient way to parameterize the matrix $\Sigma$ in 
the representation of SU(4)/Sp(4) is based on its SU(4) transformation
properties \cite{Peskin}:
\beq
\Sigma=U\overline{\Sigma}U^{\rm T}=U^2\overline{\Sigma} \ .
\label{decompose}
\eeq
Here the second equality is imposed by the algebra of the
group \cite{Peskin}. The special unitary 4$\times4$ matrix
\beq
U=e^{i\frac{\Pi}{2F}}
\eeq
contains the fluctuating fields $\Pi$.
At $\mu=0$ the Lagrangian is minimized by $\Sigma=\Sigma_M$,
aligned along the quark mass term $m{\cal M}$:
\beq
\Sigma_M\equiv\left(\begin{array}{cccc} 0 & 0 & -1 & 0 \\  0 & 0 & 0 & -1 \\
    1 & 0 & 0 & 0  \\  0 & 1 & 0 & 0  \end{array}\right) \ .
\label{Sigma_c}
\eeq
With this choice the hermitian field $\Pi$ is restricted by the manifold to \cite{Peskin}
\beq
\Pi\equiv\frac{1}{\sqrt{2}}\left(\begin{array}{cccc} 
\tilde{p} & p & 0 & q \\
p^* & -\tilde{p} & -q & 0 \\
0 & -q^* & \tilde{p} & p^* \\
q^* & 0 & p & -\tilde{p}   
 \end{array}\right) \ .
\label{Pi}
\eeq
Where $\tilde{p}$ is real and $p$ and $q$ are complex numbers. 
%We shall return to
%the choice for $\overline{\Sigma}$, for now it is sufficient to notice 
%that this choice is not unique. We say that the choice determines a
%direction.
The fields 
$\tilde{p}$, $p$, $p^*$ are neutral and charged pions and
$q$, $q^*$ are diquark/antidiquarks.
The $\overline\Sigma$ will change as a function of $\mu$ and we
shall discuss the parameterization around a generic minimum
below.

The purpose of carrying out an expansion of ${\cal L}_{\rm eff}$ in the 
$\Pi$-fields about the minimum $\overline{\Sigma}$ is twofold. First, in
order to justify that the ansatz (\ref{ansatz}) is a minimum, we need to prove
that the linear terms in $\Pi$ vanish. Second, the expansion of ${\cal
  L}_{\rm eff}$ allows us to determine the  dispersion relations for the low
energy excitations.

\noi
\subsection{Linear Terms}

To expand ${\cal L}_{\rm eff}$ around $\overline{\Sigma}$ defined in
(\ref{ansatz}) we need to choose a parameterization of the
$\Pi$-field which can be used for all values of $\al$ and $\eta$. For $\al=0$ 
we have already given a parameterization in (\ref{Pi}). For a given pair
$(\al>0,\eta)$ this parameterization of $\Sigma$ is no longer valid. 
We must expand about the rotated $\overline{\Sigma}$ according to
(\ref{decompose}). However, as it is useful not to change the parameterization
(\ref{Pi}) with $\al$ and $\eta$, we extract the rotation from $U$ and write
it explicitly. To do this we first note that we can write the ansatz (\ref{ansatz}) as
\beq
\overline{\Sigma}(\al,\eta)=e^{-\al
  (\Sigma_B\cos\eta+\Sigma_I\sin\eta)\Sigma_M}\Sigma_M\equiv
V^2_{(\al,\eta)}\Sigma_M\equiv V_{(\al,\eta)}\Sigma_MV^{\rm T}_{(\al,\eta)} \ .
\eeq
Referring to (\ref{decompose}) we see that
\beqn
\Sigma(\al,\eta) & \equiv &
U(\al,\eta)\overline{\Sigma}(\al,\eta)U^{\rm T}(\al,\eta) \nn \\
 &  = & V_{(\al,\eta)} 
U(\al=0,\eta)V^\dagger_{(\al,\eta)}V_{(\al,\eta)}\Sigma_MV^{\rm
  T}_{(\al,\eta)}{V^\dagger}^{\rm T}_{(\al,\eta)} U^{\rm T}(\al=0,\eta)V^{\rm T}_{(\al,\eta)}
\nn \\
 &= & V_{(\al,\eta)}\Sigma_M U^2(\al=0,\eta)V^{T}_{(\al,\eta)} \ .
\eeqn
Inserting this in ${\cal L}_{\rm eff}$ (\ref{L}) we find that the form of
the Lagrangian does not change provided that we substitute rotated values of
$\cal{M}$ and $\mu_BB+\mu_II$ 
\beqn
\label{rotMBI}
V_{(\al,\eta)}^T {\cal M}V_{(\al,\eta)} & = & -\Sigma_M \cos\al
-(\Sigma_B\cos\eta+\Sigma_I\sin\eta)\sin\al \nn \\
V_{(\al,\eta)}^\dagger (\mu_B B + \mu_I I)V_{(\al,\eta)} & = &
\left(\mu_B B\cos\al-\mu_B B
  \Sigma_B\Sigma_M\sin\al+\mu_I I\right)\cos\eta \\
&  & +
\left(\mu_I I\cos\al-\mu_I I
  \Sigma_I\Sigma_M\sin\al+\mu_B B\right)\sin\eta \ . \nn 
\eeqn
Using this and expanding $U=\exp(i\Pi/2F)$ to $2^{\rm nd}$ order in $\Pi$ we
find that the terms liner in $\Pi$ vanish. (The cancellation takes
place due to the equations (\ref{al},\ref{y}) arising from the constraint $\d_\al {\cal L}_{\rm
  st}(\overline{\Sigma})=0$.) From this we conclude that the ansatz
(\ref{ansatz}) for the minimum is at least a local extremum of the theory. As
we now turn to focus on the terms quadratic in $\Pi$ we shall find that this 
local extremum is indeed a minimum.

\noi
\subsection{Dispersion Relations}

The quadratic terms in $\Pi$ of the expansion around $\overline{\Sigma}$
determine the dispersion relations of the low energy excitations. For
$\mu_B>\mu_I$ the terms in ${\cal L}_{\rm eff}(\Sigma)$ at second order in
$\Pi$ are
\beqn
\label{qI}
{\cal L}_{\rm eff}^{(2)}(\Sigma) & = &
\frac{1}{2}\Tr\left\{
\left(\d_\nu\Pi
-\mu_I[I,\Pi]\delta_{\nu0}
-\mu_B\cos\al[B,\Pi]\delta_{\nu0}\right)^2\right.
\nn\\
& & 
\left. +\mu_B^2[B\Sigma_B\Sigma_M,\Pi]^2\sin^2\al
+m_\pi^2\Pi^2\cos\al\right\}
\ .
\eeqn

The corresponding expression for $\mu_B<\mu_I$ is obtained from the above by
switching $B$ and $I$ symbols.
Recall that we
have kept the parameterization (\ref{Pi}) at the cost of having rotated
$(\mu_BB+\mu_II)$ and ${\cal M}$, see (\ref{rotMBI}). 
Using this parameterization of
$\Pi$ we rewrite the above equation as
\beqn
{\cal L}_{\rm eff}^{(2)}(\Sigma) & = &  
(\d_\nu \tilde{p})^2
+(\d_\nu p - \mu_I \delta_{\nu0}p)(\d_\nu p^*+ \mu_I \delta_{\nu0}p^*)
+(\d_\nu q - \mu_B\delta_{\nu0} q\cos\al)(\d_\nu q^* + \mu_B\delta_{\nu0} q^*\cos\al) \nn \\
&&
+{\mu_B^2\over4}((q+q^*)^2+4(pp^*+\tilde{p}^2))\sin^2\al
+m_\pi^2(\tilde{p}^2+pp^*+qq^*)\cos\al \ .
\label{qII}
\eeqn
This allows us to read off the dispersion relations for each of the 5
modes.

In the diquark condensation phase $\alpha\ne0$ the $q$ and $q^*$ 
fields mix. The dispersion relations are 
obtained by solving for $E=ip_0$: % \cite{kogut2}
\beq
\det\left(\begin{array}{cc} \mu_B^2\sin^2(\al)/2 & \begin{array}{c}
    -(E-\mu_B\cos\al)^2+{\bf p}^2\\
+\mu_B^2\sin^2(\al)/2+m_\pi^2\cos\al \end{array} \\ 
 & \\ 
 \begin{array}{c} -(E+\mu_B\cos\al)^2+{\bf p}^2 \\
+\mu_B^2\sin^2(\al)/2+m_\pi^2\cos\al \end{array} & \mu_B^2\sin^2(\al)/2 
\end{array}\right) = 0 \ .
\eeq
The mixing between $q$ and $q^*$ implies that two of the low energy
excitations are linear combinations of diquark and anti-diquark
modes. 
We denote them $\tilde{q}$ and $\tilde{q}^*$.
As these diquark/anti-diquarks carry no isospin 
it is not surprising
that the dispersion relations are the same as found in
\cite{kogut2}. In the normal phase ($\al=0$)
\beqn
q^* \hspace{2cm} E& = &  \sqrt{{\bf p}^2+m_\pi^2}+\mu_B\nn \\
q \hspace{2cm} E& = &  \sqrt{{\bf p}^2+m_\pi^2}-\mu_B\ .
\label{disqal=0}
\eeqn   
while for $\al\not=0$
\beqn
\tilde{q}^* \hspace{2cm} E^2 & = &  {\bf p}^2+\mu_B^2(1+3\cos^2\al)/2
+\mu_B\sqrt{\mu_B^2(1+3\cos^2\al)^2/4+4{\bf p}^2\cos^2\al} \nn \\
\tilde{q} \hspace{2cm} E^2 & = &
{\bf p}^2+\mu_B^2(1+3\cos^2\al)/2
-\mu_B\sqrt{\mu_B^2(1+3\cos^2\al)^2/4+4{\bf p}^2\cos^2\al} 
\ .
\label{disqal}
\eeqn

\noindent
Note that one mode is massless for $\mu_B>m_\pi$.

\begin{figure}[t]
\vspace{-4cm}\

  \unitlength1.0cm
  \begin{center}
 \begin{picture}(3.0,2.0)
  \put(-7.0,-11.0){
  \epsfysize=6.5cm
  \epsfbox[110 -180 390 100]{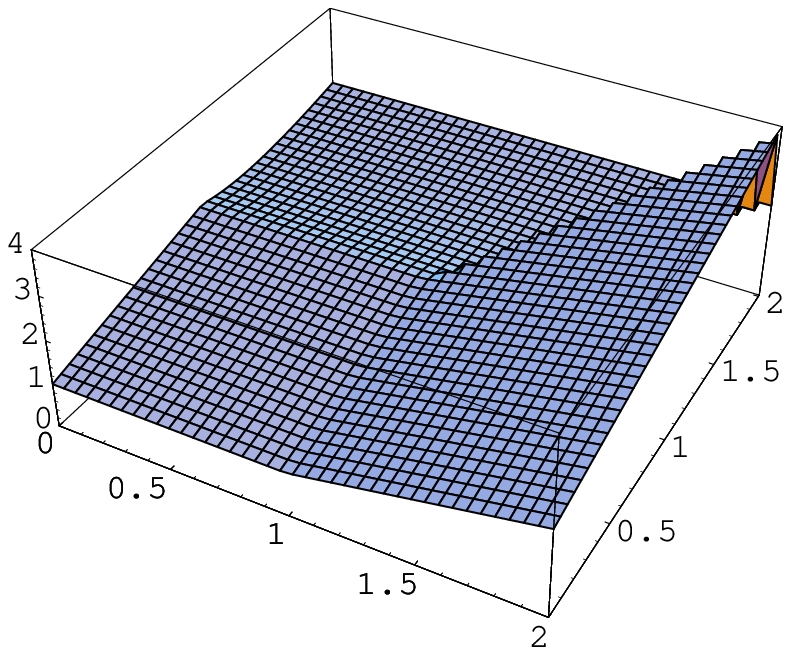}}
  \put(-5.0,-6.5){\bf\large $\frac{\mu_B}{m_\pi}$}
  \put(-2.,-5.2){\bf\large $\frac{\mu_I}{m_\pi}$}
  \put(-7.3,-3.3){\bf\large $\frac{m_{\pi^+}}{m_\pi}$}
  \end{picture}
  \end{center}

\hfill 
  \unitlength1.0cm
  \begin{center}
  \begin{picture}(3.0,2.0)
  \put(-0.8,-7.5){
  \epsfysize=6.5cm
  \epsfbox[110 -180 390 100]{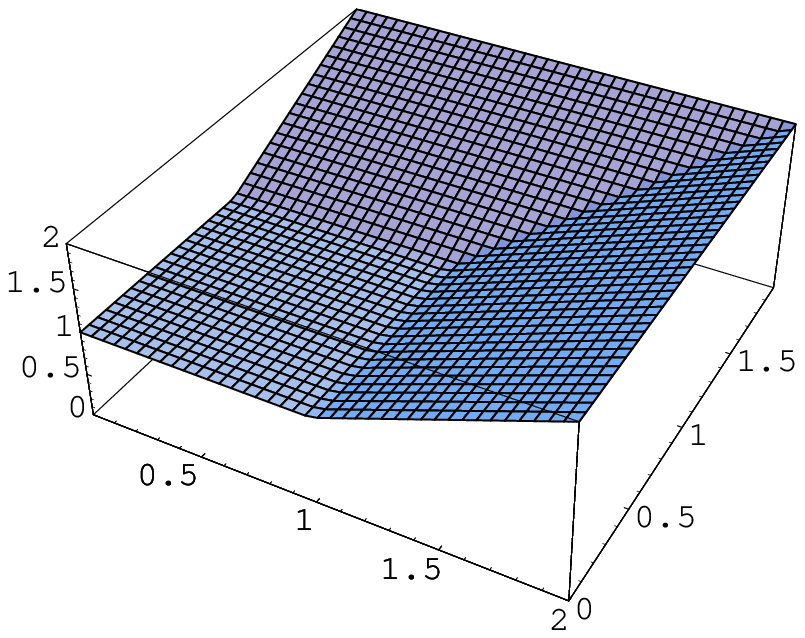}}
  \put(1.3,-3.2){\bf\large $\frac{\mu_B}{m_\pi}$}
  \put(4.,-2.){\bf\large $\frac{\mu_I}{m_\pi}$}
  \put(-1.,0.){\bf\large $\frac{m_{\pi^0}}{m_\pi}$}
  \end{picture}
  \end{center}

\hfill 
  \unitlength1.0cm
  \begin{center}
  \begin{picture}(5.0,2.0)
  \put(6.0,-4.5){
  \epsfysize=6.5cm
  \epsfbox[110 -180 390 100]{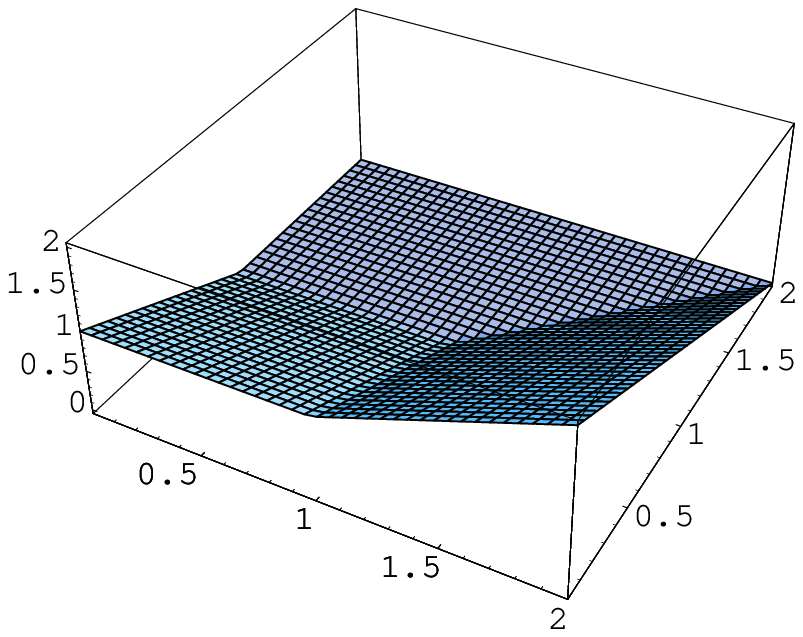}}
  \put(8.1,-0.3){\bf\large $\frac{\mu_B}{m_\pi}$}
  \put(10.7,1.){\bf\large $\frac{\mu_I}{m_\pi}$}
  \put(5.7,3.){\bf\large $\frac{m_{\pi^-}}{m_\pi}$}
  \end{picture}
  \end{center}

\caption{\label{muBmuI-figIV} $\beta_{\rm D}=1$, $N_f=2$: The ratio masses of
  the pion modes to 
  $m_\pi$ as a function of $\mu_B$ and $\mu_I$. From the left $m_{\pi^+}$, $m_{\pi^0}$,
  and $m_{\pi^-}$. Note that the mass of the $\pi^-$ excitation vanishes at
  $\mu_B=\mu_I>m_\pi$. The masses of the $\tilde{q}^*$ and the $\tilde{q}$ excitations
  are the mirrors in the $\mu_B=\mu_I$-plane of $m_\pi^+$ and $m_\pi^-$ respectively.}  
\end{figure}

There is no mixing in the $\tilde{p}$, $p$, $p^*$ sector. This implies that
we can interpret $\tilde{p}$, $p$, and $p^*$ as $\pi^0$, $\pi^+$ and $\pi^-$
and directly read off the dispersion relations: For $\al=0$ 
\beqn
\pi^0 \hspace{2cm}  E & = & \sqrt{{\bf p}^2+m_\pi^2} \nn \\
\pi^+  \hspace{2cm} E & = & \sqrt{{\bf p}^2+m_\pi^2} + \mu_I \nn \\
\pi^- \hspace{2cm}  E & = & \sqrt{{\bf p}^2+m_\pi^2} - \mu_I 
\label{dispal=0}
\eeqn
while in the diquark condensation phase ($\al\not=0$)
\beqn
\pi^0 \hspace{2cm} E & = &  \sqrt{{\bf p}^2+\mu_B^2} \nn \\
\pi^+ \hspace{2cm} E & = &  \sqrt{{\bf p}^2+\mu_B^2}  + \mu_I  \nn \\  
\pi^- \hspace{2cm}E & = &  \sqrt{{\bf p}^2+\mu_B^2}  - \mu_I \ .
\label{dispal}
\eeqn
As one might expect we find that the $\pi^+$ and $\pi^-$ modes couple to
$\mu_I$. For $\mu_B=\mu_I>m_\pi$ the $\pi^-$ becomes massless. 
For plots of the masses see figure  \ref{muBmuI-figIV}.

The dispersion relations for $\mu_B<\mu_I$ in the first quadrant are
mirrors of the ones above. To obtain them one needs only to switch
$\mu_I\leftrightarrow\mu_B$  in (\ref{disqal=0})-(\ref{dispal})
and charged pions with diquarks/anti-diquarks: $p\leftrightarrow q$.

Finally, since the masses of the excitations are positive the value of ${\cal
  L}_{\rm eff}(\overline{\Sigma})$ with $\overline{\Sigma}$ given in
(\ref{ansatz}) is not only an extremum as shown in the previous section,
but is indeed a minimum. We
assume that this minimum is global.

\section{Positivity and  $N_f=4$ phase diagram}
\label{sec:nf=4}

Testing these results in lattice Monte Carlo simulations
requires care. The fermion determinant, though always real in 2-color QCD,
is not positive definite if both $\mu_B$ and $\mu_I$
are non-zero. This can be seen by factorizing the determinant
into a product of the determinants for $u$ and $d$ quarks
with chemical potentials $\mu_{u,d}=\mu_B\pm\mu_I$.
Each of these determinants is real but not positive definite
if $\mu_{u,d}\ne0$. So, unless $\mu_u=\pm\mu_d$, their
product is not positive definite. Doubling the number
of quarks to save positivity
will take us to the $N_f=4$ theory, with a significantly
different phase diagram, which we shall now discuss.
\footnote{We assume that the chiral symmetry is still
broken at $N_f=4$. 
%It is believed that $N_f=4$ is close, from below, to the unknown critical value of $N_f$ above which the chiral symmetry is restored in 2-color QCD  \cite{ask edward?}. 
The results are trivially extended to 2$N$ $u$-quarks and
2$N$ $d$-quarks if chiral symmetry is assumed to be spontaneously broken for
that number of flavours. Likewise the results on adjoint QCD presented in the 
appendix may be extended to $N$ $u$-quarks and
$N$ $d$-quarks.}

\begin{figure}
\epsfxsize=3in
\centerline{\epsfbox{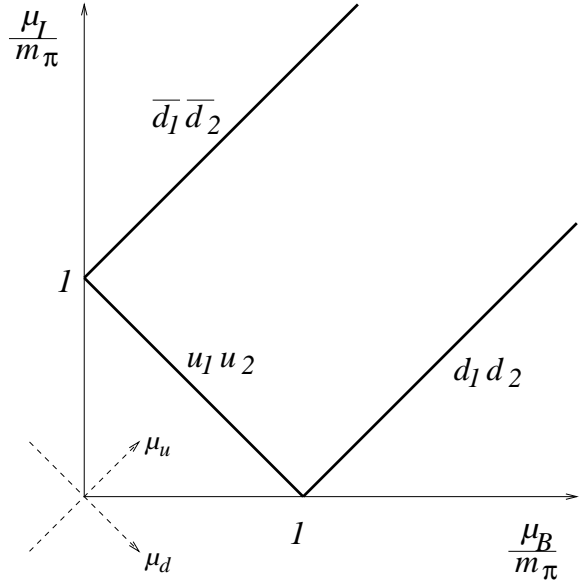}}
\caption[]{Phase diagram of the $N_f=4$ 2-color QCD, determined
by a qualitative argument of section \ref{sec:nf=4}. Solid lines
are (second order) phase transitions where certain diquark/antidiquark
condensates, indicated on the appropriate side of each line,
appear. Subscripts refer to flavour.
Dashed lines show the direction of $\mu_u$ and $\mu_d$ axes.}
\label{fig:nf=4}
\end{figure}

One can understand the phase diagram in this $N_f=4$ theory (two
{\it up}
quarks $u_{1,2}$ and two {\it down} quarks $d_{1,2}$) qualitatively using
the arguments of sect. \ref{sec:qualitative}. The most important
difference is that there are now diquarks which carry isospin: 
there is an isospin triplet ($u_1u_2$, $d_1d_2$, $u_1d_2+d_1u_2$) 
and 3 isosinglets.  
As $\mu_B$ increases, the masses of all 6 diquarks are decreasing
as $m_\pi-\mu_B$ \cite{kogut2}. Among them is $u_1u_2$,
which becomes the lightest particle with positive isospin.
Therefore, for a given $\mu_B<m_\pi$, the isospin charge condensation occurs
at $\mu_I=m_\pi-\mu_B$ (instead of the
horizontal line $\mu_I=m_\pi$ as in $N_f=2$) -- compare
figs. \ref{muBmuI-figI} and \ref{fig:nf=4}. At $\mu_B>m_\pi$
the diquarks (including the ones carrying isospin) are massless 
and the isospin is condensed already at $\mu_I=0$. By the
$\mu_B\leftrightarrow\mu_I$ reflection symmetry, the baryon charge
condensation for $\mu_I>m_\pi$ sets in at $\mu_B=0$. The region of
the phase diagram to the right of (or above) the line
$\mu_I=m_\pi-\mu_B$ contains a condensate of $u_1u_2$, which is
favoured by both $\mu_B$ and $\mu_I$. 

To complete the phase diagram in fig.\ref{fig:nf=4} note that
the line  $\mu_I=m_\pi-\mu_B$ where $u_1u_2$ diquark condensation occurs
continues also into the negative $\mu_I$ halfplane. Because
the theory is invariant under $\mu_I\to-\mu_I$, $d\leftrightarrow u$
substitution, a line $m_I=\mu_B-m_\pi$ must exist in the
positive $\mu_B\mu_I$ quadrangle (see fig.\ref{fig:nf=4}).
Below this line the condensate of diquarks $d_1d_2$, 
carrying isospin $-1$, appears. By $\mu_B\leftrightarrow\mu_I$,
$d\leftrightarrow\bar d$ reflection we obtain another line,
$\mu_I=\mu_B+m_\pi$ above which the condensate of $\bar d_1\bar d_2$
appears.

An easier way to understand the phase diagram of fig.\ref{fig:nf=4}
is to consider $u$ and $d$ quarks separately. Rotating
by $45^\circ$ and using $\mu_u$ and $\mu_d$ axes one sees
that the diagram is completely symmetric with respect to $u$
and $d$ quarks, with condensation of $u_1u_2$ and $d_1d_2$ diquarks
or antidiquarks occurring independently once $|\mu_u|$ or $|\mu_d|$
exceed the vacuum mass $m_\pi$.

We have deferred the rigorous derivation of this phase diagram
to future work. However, the phase diagram of another theory:
2- (or any-) color QCD with $N_f=2$ {\em adjoint} 
quarks (Dyson  index $\beta_{\rm D}=4$)
is derived in the Appendix. This theory is similar to $N_f=4$ 
($\beta_{\rm D}=1$), considered in this section,
in that it also has diquarks with non-zero isospin: $uu$ and $dd$.
The phase diagrams of these two theories are similar.

\section{Large $\mu$ and FFLO}
\label{sec:FFLO}

The phase diagram we obtained using chiral perturbation theory is
valid only as long as the chemical potentials remain small compared to
the chiral scale, or $m_\rho$ which is the mass of the lightest
hadron not included in the chiral Lagrangian.  We can, however, study
an opposite regime of very large $\mu$, when quarks are asymptotically
free.

\begin{figure}
\epsfxsize=3in
\centerline{\epsfbox{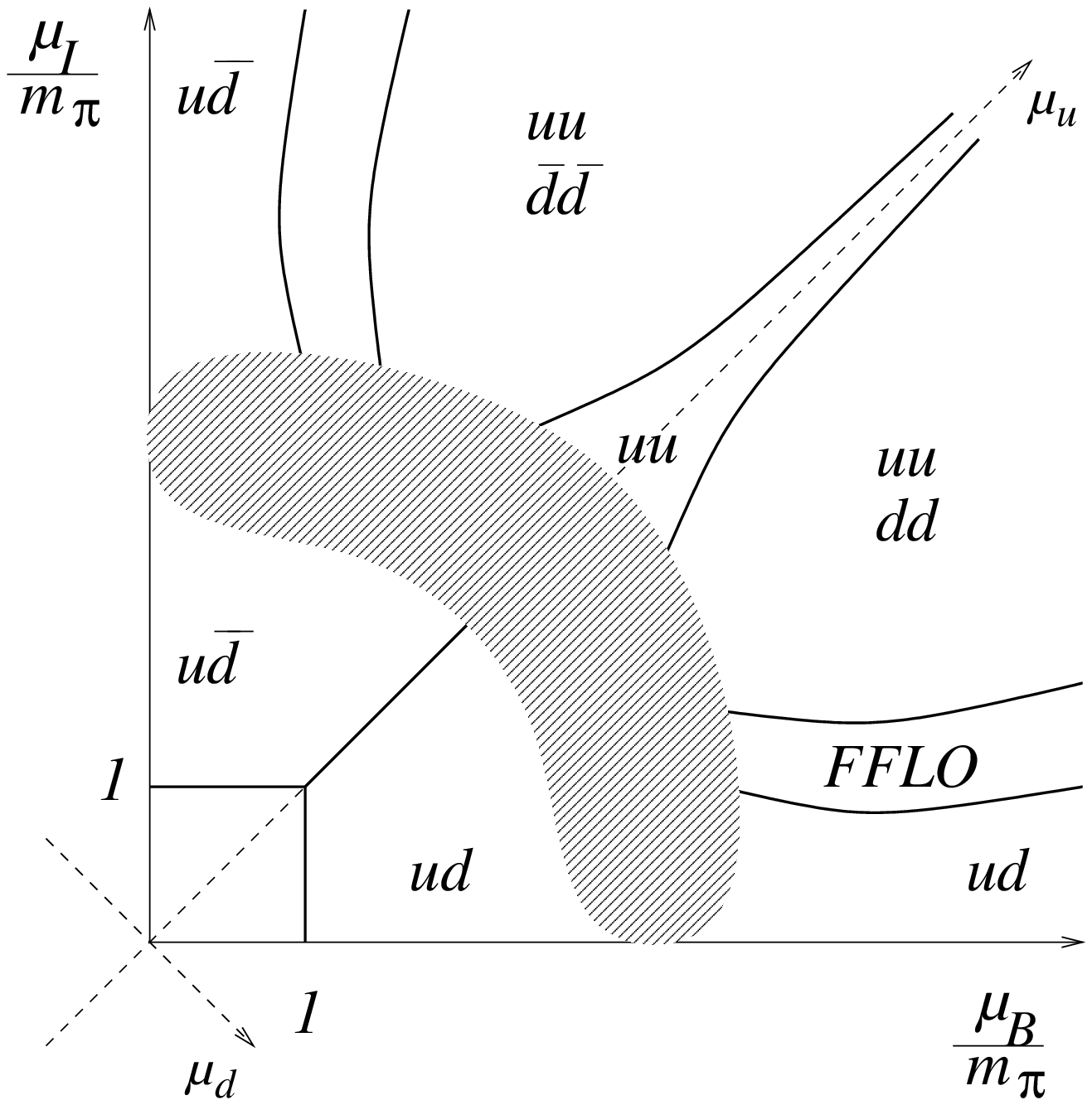}}
\hfill
\centerline{\epsfbox{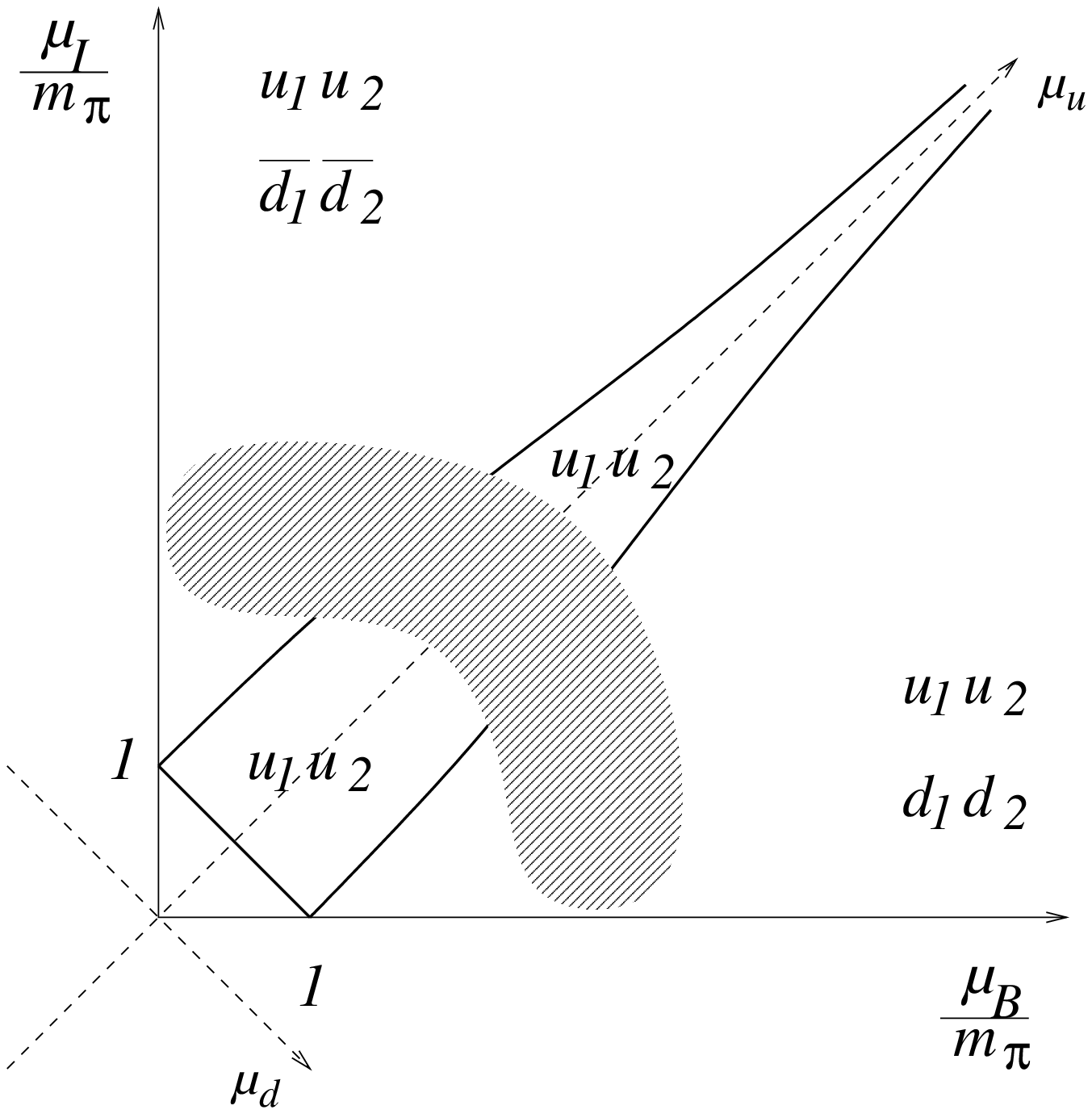}}
\caption[]{
Two colour QCD at small and large chemical potentials: Upper figure shows
$N_f=2$ and lower figure displays $N_f=4$.
Solid lines are phase transitions and gray areas illustrate regions of phase
space which remain undetermined. The lines above the clouds are drawn out of
scale. Dashed lines show the direction of $\mu_u$ and $\mu_d$ axes.}
\label{mularge}
\end{figure}

Let us first consider $N_f=2$.  On the $\mu_B$ axis, $\mu_I=0$, the
system is a Fermi liquid of $u$ and $d$ quarks with equal chemical
potentials. The ground state is a superfluid state with nonzero
$\<ud\>$ condensate, which has the same quantum number as the diquark
condensate considered previously in the framework of the chiral
perturbation theory.  Thus, it is natural to assume no phase
transition on the $\mu_B$ axis.  Analogously, the ground state at
large $\mu_I$ contains $u\bar d$ Cooper pairs, since $u$ and $\bar d$
have the same Fermi energy; and there is likely no phase transition
separating this phase from the low-$\mu_I$ phase of pion condensate.
In both cases of large $\mu_B$ and $\mu_I$ all fermions acquire a gap
$\Delta\sim\mu g^{-5} e^{-c/g}$ 
\cite{S}, here $c=3\pi^2/\sqrt{2}$ and $g$ is the (small) gauge coupling. The 
proportionality factor has been calculated in \cite{BLR}.

When $\mu_B$ and $\mu_I$ are nonvanishing, the chemical
potential for the $u$ and $d$ quarks have different magnitudes.
Generically, $\<ud\>$ and $\<u\bar d\>$ condensates cannot be formed
unless the mismatch of the Fermi momenta is small.  Thus, except for
small regions near the $\mu_B$ and $\mu_I$ axes, in most of the
$(\mu_B,\mu_I)$ plane the only condensates that can be formed are
$\<uu\>$ and $\<dd\>$ (or $\<\bar d\bar d\>$.)  These Cooper pairs are
color antisymmetric and flavor symmetric, and must hence carry spin or
orbital moment and break rotational symmetry.

The region near the diagonal $\mu_B=\mu_I$ requires special
consideration.  On the diagonal $\mu_d=0$ and no $d$-quarks are
present, while the $u$ quarks are paired.  Below the scale of the BCS
gap for the $u$ quark one should expect the $u$ quark to decouple completely.
The ($E\ll \Delta$) physics on the diagonal is thus $N_f=1$ QCD
 at zero chemical potential.  The theory is confining \cite{RSS} and one expects a gap
 for the lightest $dd$ baryon.  This baryon cannot be generated unless
$\mu_d$ is larger than this mass gap.  Therefore, there is a strip
along the diagonal where no $\<dd\>$ or $\<\bar d\bar d\>$ condensates
are present.

Consider now the regions near the axes, say, the $\mu_B$ one.  When
$\mu_I$ is small no isospin charge is generated.  One expects two
phase transitions \cite{FFLO,ABR}, at $\mu_I\approx 0.71\Delta$ and 
$0.75\Delta$.  In
the narrow window the ground state is the FFLO phase with spatial
varying diquark condensate.  Similarly, there is a strip of another
FFLO phase near the $\mu_I$ axis. We note that the evaluation in 
\cite{ABR} applies equally well to $N_c=3$ and $N_c=2$ and
support the occurrence of FFLO in two-colour two-flavour QCD.

We summarize what is said above in Fig. \ref{mularge}.  Since we can solve
our model analytically only in the two extreme limits, we have no
information about how the lines are connected in the region of
intermediate chemical potentials.  Unfortunately, two colour QCD has a sign
problem at $N_f=2$, so it is not clear whether one can find out about
it on the lattice.

For $N_f=4$, it is more convenient to work in the variables $\mu_u$ and
$\mu_d$.  When $|\mu_u|\ne|\mu_d|$, the preferable pairing is $uu$ and
$dd$ or $\bar d\bar d$.  When $\mu_d=0$, the ($E\ll\Delta$) theory reduces to
vacuum QCD with $N_f=2$ which has confinement and a mass gap for the
$d_1 d_2$ diquark.  When $\mu_d$ is less than $m_{dd}$, there should
not be any $d$ quarks in the system.  Thus, the phase diagram should
resemble fig. \ref{mularge}, where all lines are second order. 
The $u_1u_2$ condensate is always
favourable and no FFLO-type transitions occur.  

An interesting
point is that the non-homogeneous FFLO phase can occur only in the
$N_f=2$ case, and is not expected to arise in the $N_f=4$ case. 
This observation is a special case of a more general relation: 
the FFLO phase in Euclidean QCD
is excluded if the measure is positive. This might be argued
along the following lines, using QCD inequalities
\cite{QCDineq}.
The measure in the QCD-like theories is real because of the additional 
symmetries of the Dirac operator, $D$. For $N_c=2$ ($\beta_{\rm D}=1$) 
at finite baryon and isospin chemical potential the additional symmetry is 
\beq
\sigma_dC\gamma_5\tau_2 D \sigma_dC\gamma_5\tau_2 =D^*  \ \ \ {\rm with} \ \
\ [\sigma_d,\sigma_3]=0  \ \ \ {\rm and } \ \ \ \sigma_d^2=1 
\label{herm-rel}
\eeq
where $C=i\gamma_0\gamma_2$, $\tau_2$ acts in colour space, and $\sigma_3$ 
in flavour space. This relation holds for $\mu_B\neq0$ and 
$\mu_I\neq0$. If the measure is positive (by doubling of flavours) the 
relation leads to a QCD-inequality for the diquarks \cite{kogut1}. Since 
$\sigma_dC\gamma_5\tau_2$ only mix flavours with identical chemical potential 
so does the favored diquark channels for $\mu_B\neq0$ and $\mu_I\neq0$.
However, a necessary condition for the FFLO phase is that the 
$ud$ diquark channel is dominant for $\mu_B\neq0$ and 
$\mu_I\neq0$ (for degenerate masses). Hence positivity excludes the FFLO 
phase.
(The argument for QCD with quarks in the adjoint representation, $\beta_{\rm
 D}=4$, is equivalent - except in this case we need to
assume that quark annihilation diagrams can be neglected.)
It suggests that a relationship might exist between the presence of a
translation-invariance-breaking phase and the (absence of) 
positivity in the corresponding Euclidean theory.

\section{Discussion and Conclusions}

\noi

One of the main obstacles
to the study of QCD at finite baryon chemical potential is the loss of
positivity of Euclidean path integral measure.
Without the positivity standard lattice approach
fails. One of the ways to gain insight into behavior of QCD at finite
density is to study QCD-like theories with pseudo-real fermion content.
Classified according to the Dyson index of the
Dirac operator, $\beta_{\rm D}$, such theories are:
$N_c=2$ QCD with fundamental quarks --- $\beta_{\rm D}=1$, and
any-$N_c$ QCD with {\em adjoint} quarks --- $\beta_{\rm D}=4$.

Such QCD-like theories are studied here and in 
\cite{kogut1,kogut2,SS}.
%For $\beta_{\rm D}=1$, $N_f=4$ and $\beta_{\rm D}=4$, 
%$N_f=2$ the determinant is positive. 
We have studied the effects
of {\em simultaneous} baryon and isospin chemical potentials in two limits:
low and high density. The low energy limit is studied by means of
effective low-energy approach -- the chiral Lagrangian.
Recent first-principle lattice studies
of two-colour QCD at finite $\mu_B$ \cite{Hands}
have confirmed predictions from the
effective theory at nonzero $\mu_B$, with $\mu_I=0$. The results of our
paper are also testable. 
For $\beta_{\rm D}=1$, $N_f=4$ and $\beta_{\rm D}=4$, 
$N_f=2$ the fermion determinant is positive for all $\mu_B$ and $\mu_I$.

The low energy effective theory approach, in principle, 
applies even if positivity of
the microscopic theory is lost. E.g., it applies to three colour QCD
($\beta_{\rm D}=2$) at non-zero baryon and isospin 
chemical potential. However, the low-energy effective theory is valid only
for energies much less than the smallest non-Goldstone 
hadron mass, and in this regime there
is no effect of the baryon chemical potential. Low energy effective QCD is,
on the contrary, 
affected if {\em different} 
chemical potentials are introduced for different quark
flavours \cite{TV,SS}. The isospin case, where two
quark chemical potentials
 of the same magnitude, but opposite sign, are introduced, was 
studied in \cite{SS}. Comparing that work with the one presented here we 
learn that QCD with $\beta_{\rm D}=1,2$, and $4$ are very similar 
on the $\mu_I$-axis. In
all cases a pion condensate forms at $\mu_I=m_\pi$. 

In the $\mu_B,\mu_I$-plane the phase diagram distinguishes $\beta_{\rm
D}=1,2$, and $4$.  For $\beta_{\rm D}=1$ the pion condensate competes
against the diquark condensate, and a novel first order phase
transition takes place at $\mu_B=\mu_I>m_\pi$; for $\beta_{\rm D}=2$
there is no $\mu_B$ dependence (for $\mu_B<m_N$); and for $\beta_{\rm
D}=4$ the $\mu_u$ and $\mu_d$ dependence separate (to lowest order in
chiral perturbation theory).

In the high-density limit we have used weak coupling perturbation
theory to study the phase diagram. We have found that theories with
positivity do not possess a phase of inhomogeneous condensation --- the
FFLO phase. For example, $N_f=2$ QCD with adjoint quarks ($\beta_{\rm D}=4$)
is positive and does not display FFLO phase. On the contrary, $N_f=2$,
2-color QCD ($\beta_{\rm D}=1$), does have regions of FFLO phase at large
$\mu_B$ and $\mu_I$ ($\mu_B\gg\mu_I$ or $\mu_I\gg\mu_B$), while there 
is no positivity.

\noi

{\sc Acknowledgements:}\\
The authors thank the DOE Institute for Nuclear Theory at the University of
Washington for its hospitality. KS thanks D. Toublan and J. Verbaarschot for 
discussions.

\appendix

\setcounter{equation}{0}
\section{Quarks in the adjoint representation}
\label{beta=4}

\noi

In this appendix we give results for QCD with quarks in the adjoint
colour representation.
This version of QCD does not have a problem with positivity. For $N_f=1$ the 
determinant is real and positive definite (see, e.g., the first reference of 
\cite{Hands}). 
QCD with quarks in the adjoint colour representation belongs to the 
universality class labelled by Dyson index $\beta_{\rm D}=4$ \cite{V}, and 
its low energy limit is also described
by chiral perturbation theory. Following the line of section \ref{SectL}
to \ref{SectClassical} we consider $N_f=2$ and determine the 
phase diagram with $m\equiv m_u=m_d$. We expect to find a phase diagram similar to that of $\beta_{\rm D}=1$, $N_f=4$ since in both cases the lightest excitation on the $\mu_B$-axis is a diquark.

\begin{figure}[t]
  \unitlength1.0cm
  \begin{center}
  \begin{picture}(3.0,2.0)
  \put(-7.0,-9.0){
  \epsfysize=8.5cm
  \epsfbox[110 -180 390 100]{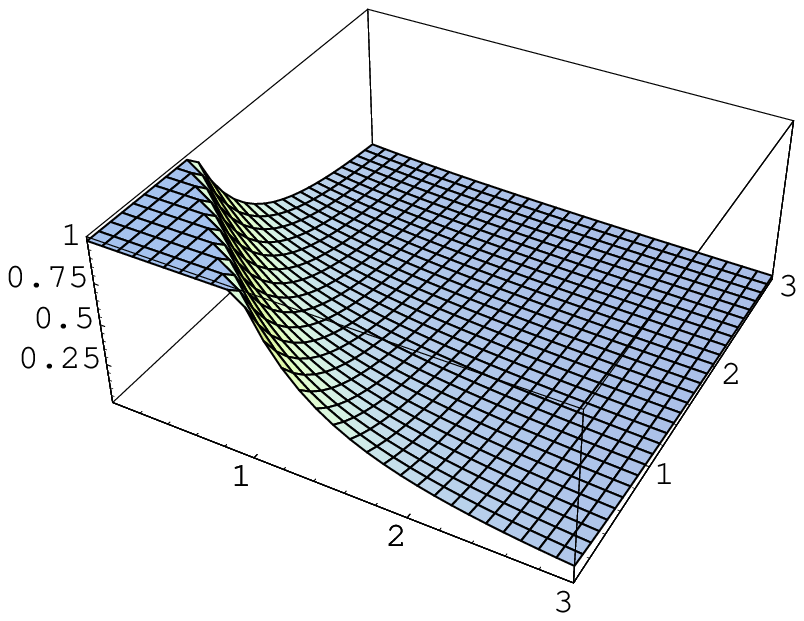}}
  \put(-4.5,-3.2){\bf\large $\frac{\mu_B}{m_\pi}$}
  \put(-1.,-2.2){\bf\large $\frac{\mu_I}{m_\pi}$}
  \put(-7.6,0.5){\bf\large $\frac{\langle\bar{\psi}\psi \rangle_u}{\langle\bar{\psi}\psi \rangle_0}$}
  \end{picture}
  \end{center}

\hfill 
  \unitlength1.0cm
  \begin{center}
  \begin{picture}(3.0,2.0)
  \put(2.0,-6.0){
  \epsfysize=8.5cm
  \epsfbox[110 -180 390 100]{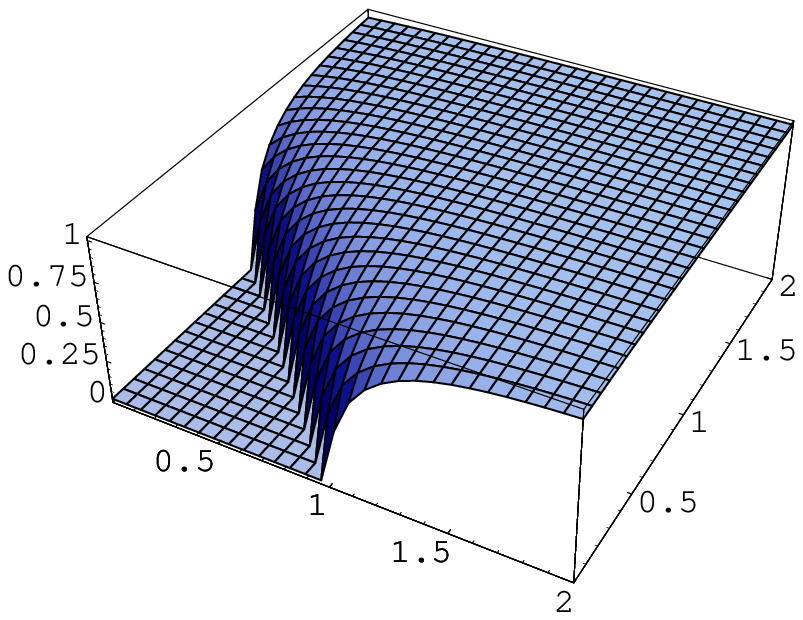}}
  \put(4.0,0.2){\bf\large $\frac{\mu_B}{m_\pi}$}
  \put(8.5,1.2){\bf\large $\frac{\mu_I}{m_\pi}$}
  \put(1.2,3.5){\bf\large $\frac{\langle\psi\psi \rangle_u}{\langle\bar{\psi}\psi \rangle_0}$}
  \end{picture}
 \end{center}

\caption{\label{muBmuI-figV}   $\beta_{\rm D}=4$: Left: Chiral condensate for $u$ in QCD with
  quarks in the adjoint colour representation. Measured in units of the
  chiral condensate at $\mu_B=\mu_I=0$. Right:  Diquark condensate for $u$.
  Plots in the $d$-sector are
  obtained from these by a rotation about $\mu_B=\mu_I=0$ by $-\pi/2$.}
\end{figure}

\begin{figure}[t]
  \unitlength1.0cm
  \begin{center}
  \begin{picture}(3.0,2.0)
  \put(-7.0,-9.0){
  \epsfysize=8.5cm
  \epsfbox[110 -180 390 100]{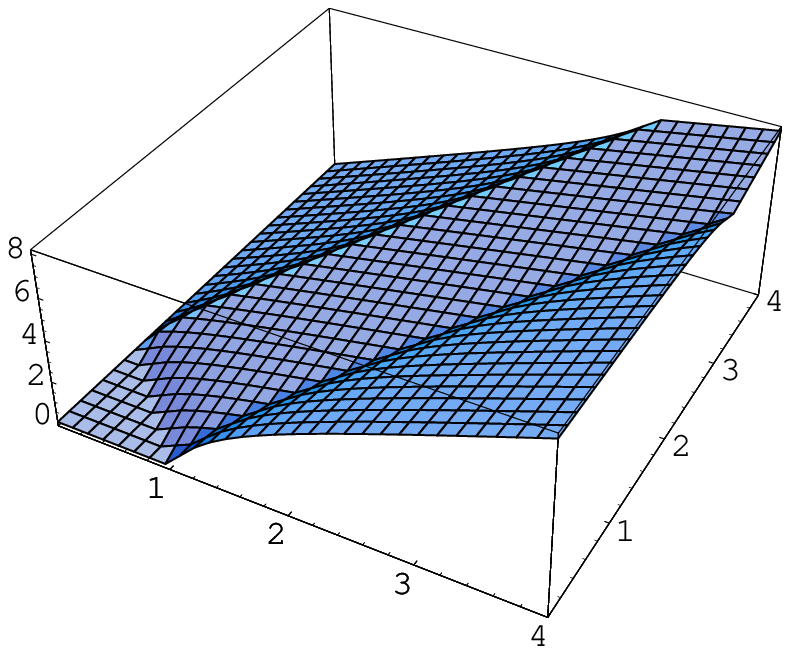}}
  \put(-4.5,-3.2){\bf\large $\frac{\mu_B}{m_\pi}$}
  \put(-1.,-2.2){\bf\large $\frac{\mu_I}{m_\pi}$}
  \put(-7.8,.5){\bf\large $\frac{n_B}{2F^2m_\pi}$}
  \end{picture}
  \end{center}

\hfill 
  \unitlength1.0cm
  \begin{center}
  \begin{picture}(3.0,2.0)
  \put(2.0,-6.0){
  \epsfysize=8.5cm
  \epsfbox[110 -180 390 100]{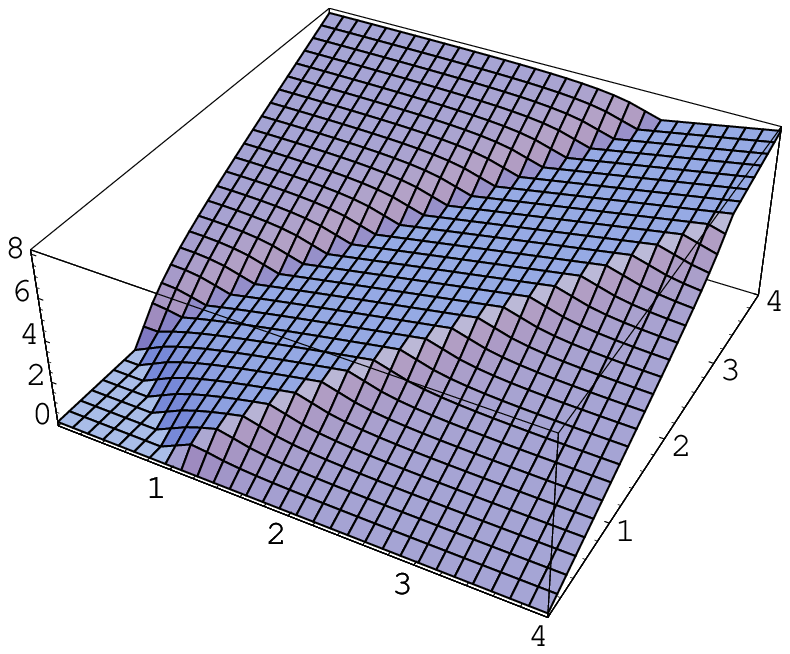}}
  \put(4.0,0.2){\bf\large $\frac{\mu_B}{m_\pi}$}
  \put(8.5,1.2){\bf\large $\frac{\mu_I}{m_\pi}$}
  \put(1.,3.5){\bf\large $\frac{n_I}{2F^2m_\pi}$}
  \end{picture}
 \end{center}

\caption{\label{muBmuI-figVI} The baryon and isospin charge densities for
  $\beta_{\rm D}=4$. Left: $n_B(\mu_B,\mu_I)$. Right  $n_I(\mu_B,\mu_I)$.}
\end{figure}

\noi
The symmetries of QCD with quarks in the adjoint colour representation allow
for a chiral goldstone manifold with $N_f(2N_f+1)-1$ degrees of freedom. The
goldstone field $\Sigma$ may in this case be represented by a special unitary
symmetric matrix \cite{Peskin}. The effective Lagrangian remains to be given
by (\ref{L}) provided that we replace ${\cal M}$ by
\beqn
\label{M4}
{\cal M}\equiv\left(\begin{array}{cccc} 0 & 0   & 1 & 0 \\  0 & 0  & 0 & 1\\
    1 & 0  & 0  & 0\\ 0 & 1  & 0  & 0\\\end{array}\right) \ . 
\eeqn
The additional degrees of freedom as compared to $\beta_{\rm D}=1$ 
$(N_f(2N_f+1)-1>N_f(2N_f-1)-1))$ allow for a
different vacuum structure. The ansatz for the vacuum field is now
\beq
\overline{\Sigma} ~ \equiv ~ \Sigma_M^{(u)}  \cos\al_u+\Sigma_u   \sin 
\al_u + \Sigma_M^{(d)} \cos\al_d  +\Sigma_d \sin 
\al_d  \ ,  
\label{ansatz4}
\eeq
where $\Sigma_M^{(u)}$ , $\Sigma_M^{(d)}$, $\Sigma_u$, and $\Sigma_d$ are defined as
\beq
\Sigma_M^{(u)} = \left(\begin{array}{cccc}
0 & 0 & 1 & 0 \\
0 & 0 & 0 & 0 \\
1 & 0 & 0 & 0 \\
0 & 0 & 0 & 0 \\ 
\end{array}\right) \ ,  \ \ 
\Sigma_M^{(d)} = \left(\begin{array}{cccc}
0 & 0 & 0 & 0 \\
0 & 0 & 0 & 1 \\
0 & 0 & 0 & 0 \\
0  & 1 & 0 & 0 \\
\end{array}\right) \ ,  \ \ 
\Sigma_u = \left(\begin{array}{cccc}
i & 0 & 0 & 0 \\
0 & 0 & 0 & 0 \\
0 & 0 & i & 0 \\
0 & 0 & 0 & 0 \\
\end{array}\right) \ , \ \ 
\Sigma_d = \left(\begin{array}{cccc}
0 & 0 & 0 & 0 \\
0 & i & 0 & 0 \\
0 & 0 & 0 & 0 \\
0 & 0 & 0 & i \\
\end{array}\right) \ .
\label{Sigma_4}
\eeq
With this ansatz the $\mu_u$ and $\mu_d$ terms do not mix at the classical
level as can be seen from the Lagrangian at the minimum
\beq
{\cal L}_{\rm eff}(\overline{\Sigma})~=~- \frac12 F^2
\left(-\cos(2\al_u)\mu_u^2-\cos(2\al_d)\mu_d^2
  +\mu_u^2+\mu_d^2+4m_\pi^2(\cos\al_u+\cos\al_d)\right) \ .
\label{LstSigma4}
\eeq
Extremizing with respect to the $\al$-directions we find:
\beqn
\al_f = 0 & {\rm if} & \frac{m_\pi^2}{\mu_f^2} > 1 \ \ , \hspace{2cm}
       f=u,d   \nn\\
\cos(\al_f) = \frac{m_\pi^2}{\mu_f^2} & {\rm if} & \frac{m_\pi^2}{\mu_f^2}
 < 1  \ .
\label{al4}
\eeqn
As for $\beta_{\rm D}=1$ we can prove that $\overline{\Sigma}$ is a local
minimum by expanding around the minimum. The proof goes along the 
lines of the $\beta_{\rm D}=1$. First we choose a representation of the 
$\Pi$ field corresponding to the direction
$\overline{\Sigma}={\cal M}$. Instead of changing this representation as 
$\overline{\Sigma}$ rotate with increasing $\mu_B$ and $\mu_I$ we again 
choose to absorb the rotation in $\mu_BB+\mu_II$ and ${\cal{M}}$. The 
analogue of Eq. (\ref{rotMBI}) is
\beqn
 V^\dagger(\al_u,\al_d)(\mu_BB+\mu_II) V(\al_u,\al_d) & = & 
 (\mu_BB+\mu_II)\left[(\Sigma_M^{(u)})^2\cos\al_u+
                      \Sigma_u\Sigma_M^{(u)\dagger}\sin\al_u\right] \nn \\
 & & 
\hspace{-1.5mm}+(\mu_BB+\mu_II)\left[(\Sigma_M^{(d)})^2\cos\al_d+
                      \Sigma_d\Sigma_M^{(d)\dagger}\sin\al_d\right] \\
 V^T(\al_u,\al_d){\cal{M}} V(\al_u,\al_d) & = & \Sigma_M^{(u)}\cos\al_u+
\Sigma_u\sin\al_u+ \Sigma_M^{(d)}\cos\al_d+
\Sigma_d\sin\al_d \ .
\eeqn
with
\beq
 V(\al_u,\al_d) ~ = ~ -i \Sigma_u e^{i\frac{\al_u}{2}\Sigma_M^{(u)}}
                      -i \Sigma_u e^{i\frac{\al_d}{2}\Sigma_M^{(d)}}  \ .
\eeq
Inserting these rotated versions of $\mu_BB+\mu_II$ and ${\cal{M}}$ into
${\cal{L}}_{\rm eff}(\overline{\Sigma})$ in Eq. (\ref{L}) we get the 
$\beta_{\rm D}=4$ effective Lagrangian at $(\al_u,\al_d)$. Expanding this to
second order in $\Pi$ one finds that the linear terms in $\Pi$ drop out
due to the extremum conditions of ${\cal{L}}_{\rm eff}(\overline{\Sigma})$ in 
Eq. (\ref{LstSigma4}) with respect to $\al_u$ and $\al_d$. Hence the ansatz 
put forward in Eq. (\ref{ansatz4}) is indeed a local extremum. Note that the 
quadratic terms in $\Pi$ mix the $u$ and $d$ sectors, i.e., the $u$ and $d$ 
sectors only separate at the classical level. As for $\beta_{\rm D}=1$ we
will evaluate the condensates and densities classically. Assuming that
 the local extremum is a global minimum we draw the following conclusions
\beq
\begin{array}{rll}
 \langle\bar{\psi}\psi\rangle_f & = & -\frac{\d {\cal L}_{\rm eff}(\overline{\Sigma})}{\d
  m_f}=2G\cos\al_f  \ , \ \ f=u,d  \nn \\ 
   \nn \\
n_B & = & -\frac{\d {\cal L}_{\rm eff}(\overline{\Sigma})}{\d \mu_B}
=2F^2[(\sin^2 \al_u+\sin^2 \al_d)\mu_B
-\mu_I\sin(\al_u+\al_d)\sin(\al_d-\al_u)]  \nn \\ 
   \nn \\
n_I & = & -\frac{\d {\cal L}_{\rm eff}(\overline{\Sigma})}{\d \mu_I} 
=2F^2[(\sin^2 \al_u+\sin^2 \al_d)\mu_I -\mu_B\sin(\al_u+\al_d)\sin(\al_d-\al_u)] \ . 
\label{condensates4}
\end{array} 
\eeq
For $\beta_{\rm D}=4$ and $N_f=2$ two of the three diquarks are flavour diagonal, 
$\psi_f^TC\gamma_5\psi_f$. Introducing diquark sources for the diagonal ones
we find
\beq
\langle\psi\psi\rangle_f =2G\sin\al_f\ , \ \ f=u,d \ .
\eeq
The flavour mixing diquark has zero vacuum expectation value outside the 
$\mu_B$ and $\mu_I$ axis. On the $\mu_B$ axis it is of course degenerate 
with the two flavour-diagonal diquark condensates. For illustrations see
figures \ref{muBmuI-figV} and \ref{muBmuI-figVI}. Finally, let us comment on 
the possibility of realizing the FFLO state
at large $\mu_B$ and small $\delta\mu_I$ (or visa versa). The theory
allows for two flavour-diagonal diquarks and one flavour-mixing diquark.
The two flavour diagonal diquark condensates are only mildly affected by
$\delta\mu_I$. The flavour mixing state exists only on the $\mu_B$ axis 
as it competes against the flavour diagonal condensates.
The FFLO state is therefore not expected to occur. This is consistent with
the general relation stated at the end of section \ref{sec:FFLO}.

\begin{thebibliography}{X}

\bibitem{BL} B.~Barrois, Nucl. Phys. {\bf B 129}, 390 (1977);
S.~Frautschi, {\em Proceedings of workshop on hadronic matter at
extreme density}, Erice 1978; B.C. Barrois, {\em
Non-perturbative effects in dense quark matter}, PhD Thesis, Caltech, 1979; 
D. Balin and A. Love, Phys. Rept. {\bf 107} (1984) 325, and
  references therein.


\bibitem{ARF}
M.~Alford, K.~Rajagopal, and F.~Wilczek,
%``QCD at finite baryon density: Nucleon droplets and color  
%superconductivity,''
Phys.\ Lett.\  {\bf B 422} (1998) 247;
%[hep-ph/9711395].
%\bibitem{CS2}
%\bibitem{Rapp:1998zu}
R.~Rapp, T.~Sch\"afer, E.~V.~Shuryak, and M.~Velkovsky,
%``Diquark Bose condensates in high density matter and instantons,''
Phys.\ Rev.\ Lett.\  {\bf 81} (1998) 53.
%[hep-ph/9711396].

\bibitem{RW} K. Rajagopal and F. Wilczek, hep-ph/0011333.

\bibitem{kogut1} J.B. Kogut, M.A. Stephanov, and D. Toublan, Phys. Lett. {\bf
    B 464} (1999) 183-191.

\bibitem{kogut2} J.B. Kogut, M.A. Stephanov, D. Toublan, J.J.M. Verbaarschot,
  and A. Zhitnitsky, Nucl. Phys. {\bf B 582} (2000) 477-513.

\bibitem{Hands}  S. Hands, I. Montvay, S. Morrison, M. Oevers, L. Scorzato,
  and J. Skullerud, Eur. Phys. J. {\bf C 17} (2000) 285-302; \\ 
R. Aloisio, V. Azcoiti, G. Di Carlo, A. Galante, A.F. Grillo, {\tt
  hep-lat/0007018} and Phys. Lett. {\bf B 493} (2000) 189-196; \\
Y. Liu, O. Miyamura, A. Nakamura, and T. Takaishi, {\tt hep-lat/0009009}; \\ 
S.J. Hands, J.B. Kogut, S.E. Morrison, and D.K. Sinclair, {\tt hep-lat/0010028}.

\bibitem{AKW}
        M.~Alford, A.~Kapustin, and F.~Wilczek,
        %``Imaginary chemical potential and finite 
        %fermion density on the lattice,''
        Phys.\ Rev.\  {\bf D 59}, 054502 (1999).
        %[hep-lat/9807039].

\bibitem{SS} D.T. Son and M.A. Stephanov, {\tt hep-ph/0005225}.

\bibitem{St96}
        M.A. Stephanov, Phys. Rev. Lett. {\bf 76} (1996) 4472;
        Nucl. Phys. Proc. Suppl. {\bf 53} (1997) 469.

\bibitem{ABR}
M. Alford, J. Bowers, and K. Rajagopal, hep-ph/0008208.


\bibitem{FFLO} P. Fulde and A. Ferrell, Phys. Rev. {\bf 134} (1964) A550;
  A.I. Larkin and Yu.N. Ovchinnikov, Sov. Phys. JETP {\bf 20} (1965) 762.  

\bibitem{SV} A. Smilga and J.J.M. Verbaarschot, Phys. Rev. {\bf D 51} (1995) 829.

\bibitem{Peskin} M. E. Peskin, Nucl. Phys. {\bf B 175} (1980) 197.

\bibitem{S} D.T. Son, Phys. Rev. {\bf D 59} (1999) 094019.  

\bibitem{BLR}W. E. Brown, J. T. Liu, and H. Ren, Phys. Rev.  {\bf D 62} (2000) 054016.
% hep-ph/9912409 
% The Transition Temperature to the Superconducting Phase of QCD at High Baryon Density

\bibitem{RSS} D. H. Rischke, D. T. Son, and M. A. Stephanov, {\tt  hep-ph/0011379}.
% Asymptotic deconfinement in high-density QCD

\bibitem{QCDineq} D. Weingarten, Phys. Rev. Lett. {\bf 51} (1983) 1830;
  S. Nussinov, Phys. Rev. Lett. {\bf 51} (1983) 2081; E. Witten,
  Phys. Rev. Lett. {\bf 51} (1983) 2351.

\bibitem{TV} D. Toublan and J.J.M. Verbaarschot, {\tt hep-th/0001110}.

\bibitem{V} J. Verbaarschot, Phys. Rev. Lett. {\bf 72} (1994) 2531-2533.


\end{thebibliography}
\end{document}